 \documentclass[aps,showpacs,amsmath,amssymbd,dvips,showkeys, preprint]{revtex4}

\usepackage{graphicx}
\usepackage{color}
\def \be  {\begin{equation}}
\def \ee  {\end{equation}}
\def \ee  {\end{equation}}
\def \bea {\begin{eqnarray}}
\def \eea {\end{eqnarray}}

\newcommand{\nn}{\nonumber}

\begin{document}

\preprint{ECTP-2013-20\hspace*{0.5cm}and\hspace*{0.5cm}WLCAPP-2013-17}

\title{Thermodynamics and higher order moments in SU(3) linear $\sigma$-model with gluonic quasi-particles}

\author{Abdel Nasser ~TAWFIK\footnote{http://atawfik. net/}}
\affiliation{Egyptian Center for Theoretical Physics (ECTP), Modern University for Technology and Information (MTI), 11571 Cairo, Egypt}
\affiliation{World Laboratory for Cosmology And Particle Physics (WLCAPP), Cairo, Egypt}

\author{Niseem MAGDY} 
\affiliation{World Laboratory for Cosmology And Particle Physics (WLCAPP), 11571 Cairo, Egypt}
\affiliation{Brookhaven National Laboratory (BNL) - Department of Physics
P.O. Box 5000, Upton, NY 11973-5000, USA}

\begin{abstract}

In framework of linear $\sigma$-model (LSM) with three quark flavors, the chiral phase-diagram at finite temperature and density is investigated. At temperatures higher than the critical temperature ($ T_c $), we added to LSM the gluonic sector from the quasi-particle model (QPM), which assumes that the interacting gluons in the strongly interacting matter, the quark-gluon plasma (QGP), are phenomenologically the same as non-interacting massive quasi-particles. The dependence of the chiral condensates of strange and non-strange quarks on temperature and chemical potential is analysed. Then, we have calculated the thermodynamics in the new approach (combination of LSM and QPM). Confronting the results with recent lattice QCD simulations shows an excellent agreement in almost all thermodynamic quantities. The first and second order moments of particle multiplicity are studied in dependence on the chemical potential but at fixed temperature and on the chemical potential but at fixed temperature. These are implemented in characterizing the large fluctuations accompanying the chiral phase-transition. The results of first and second order moments are compared with the SU(3) Polyakov linear $\sigma$-model (PLSM). Also, the resulting phase-diagrams deduced in PLSM and LSM$+$QPM are compared with each other. 
\end{abstract}


\maketitle

\section{Introduction}
\label{Introduction}

The interactions between the basic building blocks (quarks, gluons, leptons and force mediators) of the visible matter in the Universe are controlled by fundamental interactions (except gravity). The quantum electrodynamics (QED) gives a very good description for the electromagnetic phenomena. The strong interaction can be described by  the quantum chromodynamics (QCD) with an asymptotic freedom \cite{J-Gross,D-Politzer} meaning that the strong coupling  becomes small when the momentum scale for the considered processes becomes large. This leads to a phase transition from hadronic matter, in which quarks and gluons  are confined  at low temperature and density to a new state-of-mater \cite{N. Cabibbo,Collins}, called  quark-gluon plasma (QGP), in which quarks and gluons  are no longer confined  at high temperature and large density \cite{Rischke:2003mt}. The theoretical and experimental studies of QGP still represent a challenge to be faced by scientists. So far, there are many heavy-ion experiments aiming to create that  phase of  matter and to study its properties for example the ones operating with the Relativistic Heavy-Ion Collider (RHIC) and the Large Hadron Collider (LHC). From theoretical point-of-view, there are - apart from QCD and its numerical simulations - many first-principle models, like perturbative Nambu-Jona-Lasinio (PNJL) model \cite{Fukushima:2003fw,Ratti:2005jh,Fukushima:2008wg}, Polyakov linear $\sigma$-model (PLSM), Polyakov quark meson model (PQM) \cite{Schaefer:2007pw,Kahara:2008yg,Schaefer:2008ax}. Also the quasi-particle model (QPM) \cite{kmpf1,qp18a,qp18b} was suggested to reproduce the lattice QCD calculations. Each of these model has strong and weak features. The compilation between different effective models was suggested \cite{Philip:2013}, for instance, an extension in NJL to include Hadron Resonance Gas (HRG). In the present paper, we implement the same compilation aiming to fully reproduce the recent lattice QCD calculations \cite{QCDL}. LSM shall be extended to include the gluonic sector of QPM. This approach drastically reduces the dependence of QPM on the free parameters. The QPM has at least three free parameters. This will be reduced to just one parameter. In light of this, we suggest a novel approach; QCD-like model, to reproduce the most recent lattice QCD calculations.

The physical motivation of the present work shall be introduced in section \ref{sec:motv}. When confronting the thermodynamic quantities calculated in the Polyakov linear $\sigma$-model (PLSM) \cite{Tawfik:2014uka,Tawfik:2014hwa} and those calculated in the linear $\sigma$-model (LSM) to the lattice QCD calculations \cite{QCDL}  at vanishing chemical potential, we find an obvious disagreement, especially above $T_c$. We introduce new configurations to LSM so that the recent lattice simulations \cite{QCDL} will be reproduced. The reasons why the lattice calculations \cite{HotQCD} and \cite{QCDL} give different thermodynamic quantities and indirectly why should we add new configurations to LSM in order to describe the recent lattice calculations, could be originated  in the differences between \cite{HotQCD} and \cite{QCDL}.  Assuming that the interacting gluons in the strongly interacting matter are phenomenologically the same as non-interacting massive quasi-particles, we add quasigluons to the LSM.

In section \ref{Model}, we introduce the various approaches: SU(3) linear $\sigma$-model (\ref{MLS}) and gluonic sector of the quasi-particle model (\ref{MQP}). The mean field approximation is outlined in section \ref{sec:mean field}. Section \ref{sec:Results} is devoted to the results. The results of light and strange quark condensates are presented in subsection \ref{subsec:condensates}. Various thermodynamic quantities shall be calculated in subsection \ref{subsec:thermo}. The number density and susceptibility as examples of the higher order moments of particle multiplicity are outlined in section \ref{sec:higher}.
 The lattice QCD usually calculates the higher moments (other thermodynamic quantities) in temperature-dependence at fixed chemical potential. Thus, comparing our results, which are given in dependence on chemical potential at fixed temperatures, with the lattice QCD would not be possible. Since, the goal of studying higher moments is mapping out the chiral phase-diagram, we suggest mapping out the chiral boundary in chemical potential direction, section \ref{sec:chiralpt}. This, on one hand, overcomes the difficulties of the direct comparison of the higher moments, but, on the other hand, does not prevent the direct comparison of the chiral phase-diagram. The conclusions and outlook shall be given in section \ref{sec:conclusion}.

\section{The SU(3) approaches}
\label{Model}

\subsection{SU(3) linear $\sigma$-model}
\label{MLS}

The Lagrangian of LSM with $N_f =3$  quark flavors and $N_c =3$ color degrees of freedom, where the quarks couple to the Polyakov loop dynamics,  was introduced  \cite{Schaefer:2008ax,Mao:2010},
\begin{eqnarray}
\mathcal{L}=\mathcal{L}_{chiral}. \label{plsm}
\end{eqnarray}
The chiral Lagrangian $\mathcal{L}_{chiral}=\mathcal{L}_q+\mathcal{L}_m$ is of $SU(3)_{l}\times SU(3)_{R}$ symmetry  \cite{Lenaghan,Schaefer:2008hk}. 
The Lagrangian with $N_f =3$ consists of two parts.  The first part is a fermionic part, Eq. (\ref{lfermion}) with a flavor-blind Yukawa coupling $g$ of quarks. The second part is the mesonic contribution, Eq. (\ref{lmeson})   
\begin{eqnarray}
\mathcal{L}_q &=& \sum_f \overline{\psi}_f(i\gamma^{\mu}
D_{\mu}-gT_a(\sigma_a+i \gamma_5 \pi_a))\psi_f, \label{lfermion} \\
\mathcal{L}_m &=&
\mathrm{Tr}(\partial_{\mu}\Phi^{\dag}\partial^{\mu}\Phi-m^2
\Phi^{\dag} \Phi)-\lambda_1 [\mathrm{Tr}(\Phi^{\dag} \Phi)]^2 \nonumber\\&& 
-\lambda_2 \mathrm{Tr}(\Phi^{\dag}
\Phi)^2+c[\mathrm{Det}(\Phi)+\mathrm{Det}(\Phi^{\dag})]
+\mathrm{Tr}[H(\Phi+\Phi^{\dag})].  \label{lmeson}
\end{eqnarray}
The summation $\sum_f$ runs over the three flavors (f=1, 2, 3 for u-, d-, s-quark ). The flavor-blind Yukawa coupling $g$ should couple the quarks to mesons \cite{blind,RefereeRef1}. The coupling of quarks to the Euclidean gauge field \cite{Polyakov:1978vu,Susskind:1979up} $A_{\mu}=\delta_{\mu 0}A_0$ is given via the covariant derivative $D_{\mu}=\partial_{\mu}-i A_{\mu}$. In Eq. (\ref{lmeson}), $\Phi$ is a complex $3 \times 3$ matrix which depends on the $\sigma_a$ and $\pi_a$ \cite{Schaefer:2008hk}, where  $\gamma^{\mu}$ are the chiral spinors, $\sigma_a$ are the scalar mesons and  $\pi_a$ are the pseudoscalar mesons. 
\begin{eqnarray}
\Phi= T_a \phi _{a} =T_a(\sigma_a+i\pi_a),\label{Phi}
\end{eqnarray}
where $T_a=\lambda_a/2$ with $a = 0, \cdots, 8$ are the nine generators of the $U(3)$ symmetry group and $\lambda_a$ are the eight Gell-Mann matrices \cite{Gell Mann:1960}. The chiral symmetry is  explicitly broken  by $H$  
 \begin{eqnarray}
H = T_a h_a.\label{H}
\end{eqnarray}
$H$ is a $3 \times 3$ matrix with nine parameters $h_a$.

When taking into consideration that the spontaneous chiral symmetry breaking takes part in vacuum state, then a finite vacuum expectation value of the field $\Phi$ and $\bar{\Phi}$ are conjectured to carry the quantum numbers of the vacuum \cite{Gasiorowicz:1969}. As a result, the diagonal components of the explicit symmetry breaking term $h_0$, $h_3$  and $h_8$ should not vanish \cite{Gasiorowicz:1969}. This leads to exact three finite condensates $\bar{\sigma_0}$, $\bar{\sigma_3}$ and $\bar{\sigma_8}$. On other hand, $\bar{\sigma_3}$ breaks the isospin symmetry  $SU(2)$  \cite{Gasiorowicz:1969}. To avoid this situation, we restrict ourselves to SU(3). This can be $N_f= 2+1$ \cite{Schaefer:2008hk} flavor symmetry breaking pattern. Correspondingly, two degenerate light (up and down) and one heavier quark flavor (strange) are assumed. Furthermore, the violation of isospin symmetry is neglected. This facilitates the choice of $h_a$ ($h_0 \neq 0$, $h_3=0$ and $h_8 \neq 0$).  Additional to these, five other parameters should be estimated, which are the squared tree level mass of the mesonic fields $m^2$, two possible quartic coupling constants $\lambda_1$  and $\lambda_2$, Yukawa coupling $g$ and a cubic coupling constant $c$. The latter models the axial $U(1)_A$ anomaly of the QCD vacuum. It is more convenient to convert the condensates $\sigma_0$ and $\sigma_8$ into a pure non-strange and strange parts. To this end, an orthogonal basis transformation from the original basis $\bar{\sigma_0}$ and $ \bar{\sigma_8}$ to the non-strange $\sigma_q$ and strange $\sigma_s$ quark flavor basis is required \cite{Kovacs:2006}.  In LSM literature, $\sigma_q$ and $\sigma_s$ are named as $\sigma_x$ and $\sigma_y$, respectively,
\bea
\label{sigms}
\left( {\begin{array}{c}
\sigma_q \\
\sigma_s
\end{array}}
\right)=\frac{1}{\sqrt{3}} 
\left({\begin{array}{cc}
\sqrt{2} & 1 \\
1 & -\sqrt{2}
\end{array}}\right) 
\left({ \begin{array}{c}
\sigma _0 \\
\sigma _8
\end{array}}
\right).
\eea

\subsection{Quasi-Particle Model (QPM)}
\label{MQP}

The quasi-particle model gives a phenomenological description treating the interacting massless quarks and gluons as non-interacting massive quasi-particles \cite{qusim1}. The corresponding degrees of freedom are treated in a similar way as the electrons in condensed matter theory \cite{paul}; the interaction with the medium provides them with a dynamic mass. Consequently, most interactions can be taken into account. The model's three free parameters should be fixed, when confronting the model with the lattice QCD calculations. We shall notice that the resulting LSM have just one parameter.  Accordingly, the pressure at finite $T$ and $\mu$ is given as
\bea 
p &=& \sum_{i=q,g}\, p_i - B(T,\mu), \\
p_i &=& \frac{g_i}{6\, \pi^2} \int_0^{\infty} dk \frac{k^4}{E_i(k)} \left[f_i^+(k)+f_i^-(k)\right],\label{Eq:pg}
\eea
where the function $B(T, \mu)$ is there to assure thermodynamic self-consistency  \cite{ref362} that $n=\partial p/\partial \mu$, $s=\partial p/\partial T$ and $\partial p/\partial m_i^2=0$ \cite{ref364}. The distribution function reads
\bea \label{Eq:fi}
f_i^{\pm}(k) &=& \frac{1}{\exp\left[\frac{E_i(k)\mp \mu}{T}\right]\pm 1},
\eea
where $\mp$ stands for fermions and bosons, respectively, and the quasi-particle dispersion relation can be approximated by the asymptotic mass shell expression near the light cone \cite{kmpf1,qp18a,qp18b},
\bea \label{Eq:Ei}
E_i^2(k) &=& k^2 + m_i^2(T,\mu)=k^2 + \Pi_i(k; T,\mu) + (x_i\, T)^2, 
\eea
where $\Pi_i(k; T,\mu)$ is the self-energy at finite $T$ and $\mu$ and $x_i^2$ is a factor taking into account the mass scaling as using in lattice QCD simulations. A suitable parametrization of $\Pi_i(k; T,\mu)$ is given by the hard thermal loop self-energies \cite{kmpf1,ref362}.  But, the running coupling in $\Pi_i(k; T,\mu)$ should then be replaced by an effective one, $G^2(k; T,\mu)$, which can be adjusted to reproduce the lattice QCD calculations \cite{qp17} and reflect the non-perturbative effects.

Then, the net quark number and entropy, respectively, are given as
\bea
n&=& \sum_{i=q,g}\, \frac{g_i}{2\, \pi^2} \int_0^{\infty} k^2\, dk  \left[f_i^+(k)-f_i^-(k)\right], \\
s&=& \sum_{i=q,g}\, \frac{g_i}{2\, \pi^2\, T} \int_0^{\infty} k^2\, dk \left\{\frac{\frac{4 k^2}{3}+m_i^2}{E_i(k)} \left[f_i^+(k)+f_i^-(k)\right] - \mu  \left[f_i^+(k)-f_i^-(k)\right]\right\}.
\eea

The lattice QCD simulations at finite $T$ and $\mu$ prefer the use of \cite{Ejiri,Gendp3b}, 
\bea
p(T,\mu)&=&p(T,\mu=0) +\Delta p(T,\mu)= p(T,\mu=0) + T^4 \sum_{j=2}^{\infty}\, c_j \left(\frac{\mu}{T}\right)^j. \label{eq:pLQCDmu}
\eea
Thus, the coefficients of Eq. (\ref{eq:pLQCDmu}) are given by the derivative at the point of vanishing chemical potential $c_j=(T^{j-4}/j!)\partial^j p/\partial \mu^j$ \cite{kmpf1}
\bea
c_2 &=& 3 \frac{n_f}{\pi^2 T^3} \int_0^{\infty} k^2 dk \frac{\exp(\omega)}{[\exp(\omega)+1]^2}, \\
c_4 &=& \frac{1}{4} \frac{n_f}{\pi^2 T^3} \int_0^{\infty} k^2 dk \frac{\exp(\omega)}{[\exp(\omega)+1]^4} \, \left\{\exp(2\omega)-4\exp(\omega)-\frac{A_2}{\omega}\left[\exp(2\omega)-1\right] +1\right\},\\
c_4 &=& \frac{3}{385} \frac{n_f}{\pi^2 T^3} \int_0^{\infty} k^2 dk \frac{\exp(\omega)}{[\exp(\omega)+1]^6} \left\{\exp(4\omega)-26\exp(3\omega)+66\exp(2\omega)-26\exp(\omega) +1 \right. \nn \\ &-& \left. \frac{10}{3}\frac{A_2}{\omega}\left[\exp(4\omega)-10\exp(3\omega)+10\exp(\omega)-1\right] \right. \nn \\
&+& \left. \frac{4}{3}\frac{A_2^2}{\omega^2}\left[\exp(4\omega)-2\exp(3\omega)-6\exp(2\omega)-2\exp(\omega)+1\right] \right.\nn \\
&+& \left. \left(\frac{5}{3}\frac{A_2^2}{\omega^3}-10T^2\frac{A_4}{\omega}\right) \left[\exp(4\omega)+2\exp(3\omega)-2\exp(\omega)-1\right]
\right\},
\eea
where the following parameters are defined at the point of vanishing chemical potential $\mu$
\bea
A_2 &=& \frac{G^2}{\pi^2} + \frac{T^2}{2} \frac{\partial^2\, G^2}{\partial \mu^2}, \\
A_4 &=& \frac{1}{\pi^2} \frac{\partial^2\, G^2}{\partial \mu^2} +  \frac{T^2}{12} \frac{\partial^4\, G^2}{\partial \mu^4}, \\
\omega &=& \frac{1}{T} \sqrt{k^2+\frac{1}{3}T^2 G^2}.
\eea
In deriving the previous expressions, flow equation \cite{ref362} was utilized 
\bea
a_{\mu} \frac{\partial\, G^2}{\partial \mu} + a_{T} \frac{\partial\, G^2}{\partial T} &=& a_{\mu T},
\eea
where the functions $a_{\mu}$, $a_{T}$ and $a_{\mu T}$ can be determined  under the condition $a_{\mu}(T,\mu=0)=a_{T}(T,\mu=0)$ \cite{kmpf1,qm16}.

The derivatives of coupling $G$ at $\mu=0$ are given as  
\bea
\left.\frac{\partial^2\, G^2}{\partial \mu^2}\right|_{\mu=0} &=& \left.\frac{1}{a_{\mu}} \left(\frac{\partial\, a_{\mu T}}{\partial \mu} - \frac{\partial\, a_{T}}{\partial \mu} \frac{\partial\, G^2}{\partial T}\right)\right|_{\mu=0}, \\
\left.\frac{\partial^4\, G^2}{\partial \mu^4}\right|_{\mu=0} &=& \left.\frac{1}{a_{\mu}} \left(\frac{\partial^3\, a_{\mu T}}{\partial \mu^3} - \frac{\partial^3\, a_{T}}{\partial^3 \mu} \frac{\partial\, G^2}{\partial T} - 3 \frac{\partial^2\, a_{\mu}}{\partial \mu^2}\frac{\partial^2\, G^2}{\partial \mu^2} \right. \right.\nn \\
&-&\left. \left.\frac{3}{a_{\mu}} \frac{\partial\, a_{T}}{\partial \mu} 
\left[
  \frac{\partial^2\, a_{\mu T}}{\partial \mu \partial T} - 
  \frac{\partial^2\, a_{T}}{\partial \mu \partial T} \frac{\partial\, G^2}{\partial T} -
  \frac{\partial\, a_{T}}{\partial \mu} \frac{\partial^2\, G^2}{\partial T^2} -
  \frac{\partial\, a_{\mu}}{\partial T} \frac{\partial^2\, G^2}{\partial \mu^2}  
\right]
\right)\right|_{\mu=0}.
\eea

As we are interested in the gluonic sector, the self-energy for the gluons reads \cite{Bluhm}
\bea \label{pi:Ei}
\Pi_g(k; T,\mu)=\left(\left[3+\dfrac{N_f}{2}\right] T^2 + \dfrac{3}{2 \pi ^2} \sum_f \mu_{f}^2\right) \dfrac{G^2}{6},
\eea
where the effective coupling $G$ at vanishing chemical potential is given as,
\bea  \label{G:Ei}
G^2(T) &=& \left\{ \begin{array}{ll}
G^2_{\text{2loop}}(T), & T\geq T_c, \\
& \\
G^2_{\text{2loop}}(T)+b\left(1-\frac{T}{T_c}\right), & T< T_c, \end{array}
\right.,
\eea
and the two-loop effective coupling $ G^2_{\text{2loop}}(T) $ is given as \cite{kmpf1}
\bea  \label{Gloop:Ei}
G^2_{\text{2loop}}(T) &=& \frac{16\, \pi^2}{\beta_0 \ln(\xi^2)} \left[1-2 \frac{\beta_1}{\beta_0^2} \frac{\ln(\ln(\xi^2))}{\ln(\xi^2)}\right],
\eea
where 
\bea
\xi &=& \lambda \frac{T-T_s}{T_c},
\eea
and $T_s$ is a regulator at $T_c$. The parameter $\lambda$ is used to adjust the scale as found in lattice QCD simulations. The $\beta$ function \cite{betaf} depends on the coupling $g$, 
\bea
\beta &=& \frac{\partial\, g}{\partial\, \ln(\Delta_{\mu})},
\eea
with $\Delta_{\mu}$ is the energy scale. The two-loop perturbation estimation for $\beta$ functions give,
\bea
\beta_0 &=& \frac{1}{3} \left(11\, n_c - 2\, n_f\right), \\
\beta_1 &=& \frac{1}{6} \left(34\, n_c^2 - 13\, n_f\, n_c + 3\, \frac{n_f}{n_c}\right).
\eea

 
\section{Motivations}
\label{sec:motv}

\begin{figure}[htb]
\centering{
\includegraphics[width=6.cm,angle=-90]{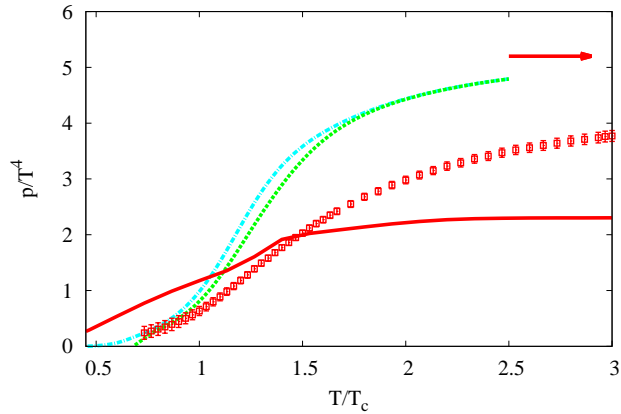}
\caption{The thermal behavior of PLSM (dash-dotted and dashed curves) \cite{Tawfik:2014uka} and lattice QCD pressure  \cite{QCDL} (empty rectangles) at vanishing chemical potential. The flatter curve (solid) represents the results from LSM (without Polyakov or quasi-particle contributions). \label{fig:prold}}
}
\end{figure}

In Fig. \ref{fig:prold}, the pressure calculated in the Polyakov linear $\sigma$-model (PLSM) \cite{Tawfik:2014uka} (dash-dotted and dashed curves) is compared with the lattice QCD calculations  \cite{QCDL} (empty rectangles) at vanishing chemical potential. The general $T$-dependence of PLSM-pressure is not absence. For instance, the pressure increases with $T$ until it gets close to the value of massless gas (Stefan-Boltzmann limit $5.2$, shown as an arrow on top right corner). The dash-dotted curve represents the results at zero-temperature value of chiral condensates, $\sigma_{x\, 0}=92.4~$ MeV, $\sigma_{y\, 0}=94.5~$ MeV and Yukawa coupling $g=6.5$. We find that increasing Yukawa coupling to $10.5$ results in the dashed curve, which practically does not differ from the solid curve. In other words, both curves fit well the lattice QCD calculations \cite{HotQCD} (circles) \cite{Tawfik:2014uka}. It is obvious that the agreement between PLSM and the recent lattice QCD \cite{QCDL} (rectangles) is not convincing, especially above $T_c$. The situation becomes worse when comparing LSM (solid curve) with the lattice results.
This is the motivation to introduce new configurations, including a gluonic potential part, to the LSM so that the most recent lattice simulations \cite{QCDL} (rectangles) are also reproduced by this QCD-like model. It is worthwhile to mention that the critical temperature is not universally constant in all these results. For instance, Refs. \cite{QCDL} and \cite{HotQCD} assume that $T_c\simeq150~$ and $195~$MeV, respectively, while PLSM suggests that $T_{\chi}=240~$MeV, Fig. \ref{fig:prold}. 

The reasons why lattice calculations \cite{HotQCD} and \cite{QCDL} give different thermodynamic quantities and indirectly why should we add new configurations to LSM in order to describe the most recent lattice calculations, can be originated  in the differences between Ref. \cite{HotQCD} and Ref. \cite{QCDL}. Therefore, a short comparison between the two sets of lattice calculations is now in order. 
\begin{enumerate}
\item Ref. \cite{QCDL} presented a full result for $2+1$ quark flavors, where all systematics are controlled, the quark masses are set to their physical values and the continuum  extrapolation is carried out. Larger lattices and a Symanzik improved gauge and a stout-link improved staggered fermion action are implemented. Depending on the exact definition of the observables, the remnant of the chiral transition is at about $T_c=150$~MeV. Extending these results, the transition temperature  was also determined for small non-vanishing baryonic chemical potentials. At high $T$, the lattice pressure is $\sim30\%$ lower than Stefan-Boltzmann limit.
\item Ref. \cite{HotQCD} used $2+1$ quark flavors with physical strange quark mass and almost physical light quark masses. The calculations have been performed with two different improved staggered fermion actions, the asqtad and p4 actions. Overall, a good agreement  is obtained between ${\cal}O(a^2)$ results from both improved staggered fermion discretization schemes, where $a$ is the lattice spacing. At high $T$, the lattice pressure is $\sim 14\%$ lower than Stefan-Boltzmann limit.
\end{enumerate}

It is now apparent that the two lattices have almost the same configurations. But, in the implemented actions lies the essential difference. The taste-changing interactions are likely through gloun exchanging with one or more components of momentum near the cutoff leading to the breaking of SU(4) vector symmetry among tastes (order $a^2$ lattice artifacts are generated). Suppressing taste-changing interactions as in {\it improved staggered fermion} or {\it asqtad actions} is frequently utilized. To achieve decreasing of the taste-violating interactions and reducing of the coupling of high-momentum gluons to the quarks, smearing the the gauge links becomes proper procedure. In {\it asqtad action}, this procedure is accomplished through replacing the gluon links in the derivatives by averages over $1$-, $3$-, $5$-, and $7$-link paths. Other actions reduce taste-changing interactions even further through more smearing. The present work and Ref. \cite{Tawfik:2014uka} seem to highlight the sensitivity of the improved staggered fermion actions and the suppression of the taste-changing interactions in the lattice QCD simulations. 

Now, we come back to the modified LSM-Lagrangian,  which can be defined according to Eqs. (\ref{plsm}) and (\ref{Eq:pg}),
\begin{eqnarray}
\mathcal{L}=\mathcal{L}_{chiral} - U_g \label{plsmg}
\end{eqnarray}
This new Lagrangian, Eq. (\ref{plsmg}), has the same symmetry as the original one, Eq. (\ref{plsm}).


\section{Mean field approximation }
\label{sec:mean field} 
In mean field approximation, we start from the partition function and then calculate the grand potential. At thermal equilibrium, the grand partition function can be defined by using a path integral over the quark, anti-quark and meson field. 
\begin{eqnarray}
\mathcal{Z}&=& \mathrm{Tr \,exp}[-(\hat{\mathcal{H}}-\sum_{f=u, d, s}
\mu_f \hat{\mathcal{N}}_f)/T] \nonumber \\ && 
= \int\prod_a \mathcal{D} \sigma_a \mathcal{D} \pi_a \int
\mathcal{D}\psi \mathcal{D} \bar{\psi} \mathrm{exp} \left[ \int_x
(\mathcal{L}+\sum_{f=u, d, s} \mu_f \bar{\psi}_f \gamma^0 \psi_f )
\right],
\end{eqnarray}
where $\int_x\equiv i \int^{1/T}_0 dt \int_V d^3x$ and $V$ is the volume of the system. $\mu_f$ is the chemical potential for quark flavors $f=(u, d, s)$. We consider symmetric quark matter and define a uniform blind chemical potential $\mu_f \equiv \mu_{u, d} = \mu_s$. Then, we evaluate the partition function in the mean field approximation  \cite{Schaefer:2008hk,blind,RefereeRef1}. We replace the meson fields by their expectation values ($\bar{\sigma_q}$ and $\bar{\sigma_s}$) in the action \cite{Mao:2010}. We can use standard methods \cite{Kapusta:2006pm} in order to calculate the integration. This gives the effective potential for the mesons. The approach introduced in the present work is as follows. The thermodynamic potential density consists of three parts:  mesonic and quasi-gluonic potentials in additional to the quark potential. 
\begin{eqnarray}
\Omega(T, \mu)=\frac{-T\, \ln (Z)}{V}=U(\sigma_q,\, \sigma_s) + U_{g}(T,\, \mu) + \Omega_{\bar{q} q}, \label{potential}
\end{eqnarray}
\begin{itemize}
\item First, the quark potential part is given as \cite{Schaefer:2008hk}
\begin{equation}
\label{eq:quark_pot}
\Omega_{\bar{q}q}(T,\mu_f) = d_q T \sum_{f=u,d,s}
\int\limits_0^\infty \! \frac{d^3 k}{(2\pi)^3} { \ln \left[1-n_{q,f}(T,\mu_f)\right] + \ln \left[1-n_{\bar{q},f}(T,\mu_f)\right] },
\end{equation}
where the occupation quark numbers read,
\begin{equation}
n_{q,f}\left(T,\mu_{f}\right)=\frac{1}{1+\exp\left[(E_{q,f}-\mu_f)/T\right]},
\end{equation}
and that for antiquarks $n_{\bar q,f}(T,\mu_{f}) \equiv n_{q,f} (T,-\mu_{f})$, respectively. The number of internal quark degrees of freedom is denoted by $d_q=2$ and $N_{c}=6$. The energies are given as
\begin{equation}
E_{q,f}= \sqrt{k^2 + m_f^2},
\end{equation}
with the quark masses $m_f$ which related to $m_q$ and $m_s$ for $u$-, $d$- and $s$-quark, respectively. 
\begin{eqnarray}
m_q &=& g \frac{\sigma_q}{2}, \label{qmass} \\
m_s &=& g \frac{\sigma_s}{\sqrt{2}} \label{sqmass},
\end{eqnarray} 
where the Yukawa coupling $g=8.3$. 

\item Second, the purely mesonic potential part reads
\begin{eqnarray}
U(\sigma_q, \sigma_s) &=& \frac{m^2}{2} \left(\sigma^2_x+\sigma^2_y\right)-h_x
\sigma_q-h_y \sigma_s-\frac{c}{2\sqrt{2}} \sigma^2_x \sigma_s
+\frac{\lambda_1}{2} \sigma^2_x \sigma^2_y + \nn \\
& & \frac{1}{8} \left(2 \lambda_1 + \lambda_2\right)\sigma^4_x + \frac{1}{4} \left(\lambda_1+\lambda_2\right)\sigma^4_y. \label{Upotio}
\end{eqnarray}
According to Eq. (\ref{newpotential}) and Fig. \ref{fig:LSMpotential}, this part will be neglected.
 
\item Third, the quasigluonic potential part

By taking into consideration gluons, this potential part can be constructed from Eqs.  (\ref{Eq:pg}), (\ref{Eq:fi}) and (\ref{Eq:Ei}), 
\bea 
U_g(T,\mu) &=& -\frac{d_g}{6\, \pi^2} \int_0^{\infty} dk \frac{k^4}{\sqrt{k^2 + m_g^2(T,\mu) + \Pi_g(k; T,\mu) + (x_g\, T)^2}} \nn \\
       & & \hspace*{1cm} \left[\frac{1}{\exp\left[\frac{\sqrt{k^2 + m_g^2(T,\mu) + \Pi_g(k; T,\mu) + (x_g\, T)^2} - \mu}{T}\right]- 1} + \right. \nn \\
       & & \hspace*{12mm}  \left. \frac{1}{\exp\left[\frac{\sqrt{k^2 + m_g^2(T,\mu) + \Pi_g(k; T,\mu) + (x_g\, T)^2} + \mu}{T}\right]- 1}\right]. \label{Eq:Ug}
\eea 
In Eq. (\ref{Eq:Ug}), the degeneracy factor $ d_g = 8$ and two parameters $ \lambda $ and $ T_s $ should be fixed in order to reproduce the lattice  QCD calculations. When adding the three potentials given in expressions (\ref{Eq:Ug}), (\ref{Upotio}) and (\ref{eq:quark_pot}), the thermodynamics and chiral phase transition can be analysed. 
\end{itemize}


\section{Results}
\label{sec:Results}
In this section, we introduce the results from this approach, Eq. (\ref{plsmg}). We introduce the  light and strange chiral condensates, thermodynamic quantities, quark number density and susceptibility and finally chiral phase-transition. Most of our results are calculated at vanishing chemical potential as we mainly want to  compare with the lattice QCD simulations, which are very reliable at this limit. The calculations at finite chemical potentials are straightforward.

\subsection{Light and strange chiral condensates}
\label{subsec:condensates}

We study  dependence of the chiral condensates $\sigma_q$ and $ \sigma_s$ on temperature and chemical potential. In order to evaluate the  $\sigma_q$ and $\sigma_s$, we minimize the thermodynamic potential, Eq. (\ref{potential}), with respect to $\sigma_q$ and $ \sigma_s$. Doing this, we obtain a set of two equations of motion
\begin{eqnarray}\label{cond1}
\frac{\partial \Omega}{\partial \sigma_q}= \left.\frac{\partial
\Omega}{\partial \sigma_s}\right|_{min} =0,
\end{eqnarray}
where the subscript $min$ means $\sigma_q=\bar{\sigma_q}, \sigma_s=\bar{\sigma_s}$, i.e. global minimum. From Eq. (\ref{cond1}) we can calculate the variation of each variable with respect to T and $\mu$ at fixed value for the other variable at the global minimum. At the global minimum, we obtain vanishing potential as a function of temperature and  chemical potential. This gives a set of points representing the behavior in thermal and dense medium. Repeating this with the other variable, we also get its thermal and dense behavior. Thus, the thermodynamics and even the chiral phase-transition turn to be accessible. 

\begin{figure}[htb]
\centering{
\includegraphics[width=7.cm,angle=-90]{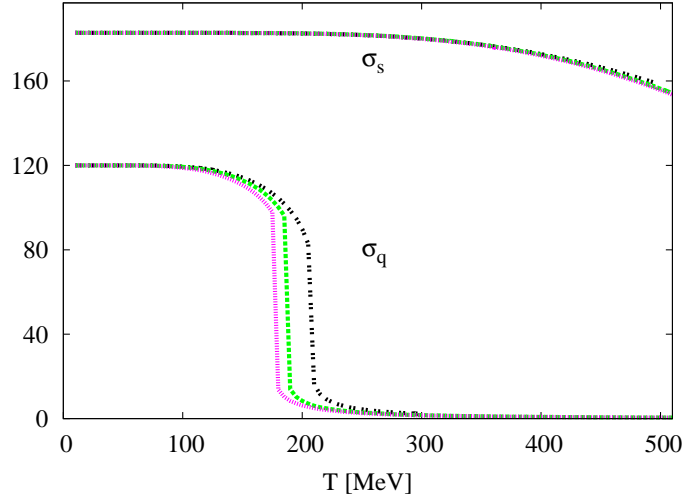}
\caption{Light chiral condensates $\sigma_q$ and $ \sigma_s$ are given as functions of temperature at different chemical potentials, $0~$MeV (double dashed curve), $50~$MeV (dotted curve) and $100~$MeV (dashed curve) \label{fig:condatMuzero} 
}
}
\end{figure}

\begin{figure}[htb]
\centering{
\includegraphics[width=7.cm,angle=-90]{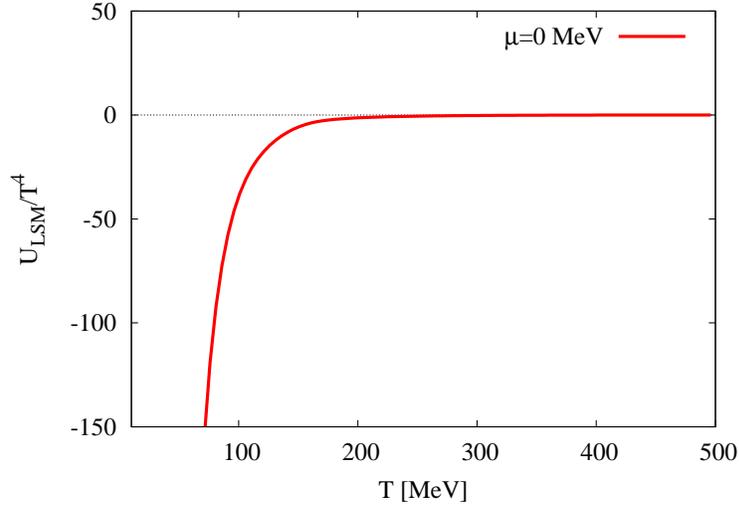}
\caption{The thermal evolution of the mesonic potential of LSM is analysed at vanishing chemical potential. Accordingly, this part of potential can be excluded, especially at high temperatures. 
\label{fig:LSMpotential} 
} }
\end{figure}

In Fig. \ref{fig:condatMuzero}, the thermal evolution of the  chiral condensates, $\sigma_q$ and $ \sigma_s$, for light and strange quark flavors is analysed at fixed chemical potentials. We notice that the values of $\sigma_q$ is less than that of $ \sigma_s$. 
We also find that $\sigma_q$ decreases faster than $\sigma_s$. The thermal behavior of the two chiral condensates is conjectured to characterize the chiral phase-transition. Below $T_c$, we find almost no change in the quark masses referring to constant condensates. With increasing $T$, the system is derived to the symmetric region, in which the quarks lose their masses.

Some remarks on $\sigma_q$ and $\sigma_s$ are in order. In Fig. \ref{fig:prold}, the differences between the critical (transition) temperatures between the LSM, PLSM and LQCD are illustrated. The transition in LSM seems to be faster than that in both PLSM and lattice QCD. In order to have a slower transition compared to the recent LQCD, we are enforced to increase the mass of the quark. Increasing the quark mass leads to an increase in the zero-value of strange and non-strange condensates.  The coupling constant should be increased as well. The consequences of all these can be seen in Fig. \ref{fig:condatMuzero}, especially in the condensates, themselves. We notice that the strange condensate is very sensitive.

Furthermore, in getting the sets of quantities representing the chiral condensates, we utilized numerical methods. Accordingly, the uncertainty is not negligible. Afterwards, we replaced all these by a mathematical expression, i.e. fitting.  Again, this method is accompanied by an additional source of uncertainty. To the best of out knowledge, there was no description about the procedure implemented in deducing the chiral condensates in any of previous works. Otherwise, we would check and accordingly cite these.

\subsection{Thermodynamic quantities}
\label{subsec:thermo}

In this section, we implement the results obtained in section \ref{subsec:condensates} in order to estimate the thermodynamics of the new approach. Also, we compare the results with the most-recent lattice QCD calculations \cite{QCDL}. To this end, we first need to prove that the behavior of the purely mesonic potential contribution, Eq. (\ref{Upotio}), should be excluded. In Fig. \ref{fig:LSMpotential}, the normalized mesonic potential contribution is given as a function of $T$ at vanishing $\mu$.  At low $T$, this part of the potential becomes infinity and vanishes at high $T$. Accordingly, the effective potential, Eq. (\ref{potential}), can be re-written as 
\begin{eqnarray}
\Omega(T, \mu) &\simeq &U_{g}(T, \mu)+\Omega_{\bar{q} q}. \label{newpotential}
\end{eqnarray}

\begin{itemize}
\item{\it The pressure. }\\
The pressure density $P$ can obtained directly from the grand potential 
 \begin{eqnarray}
p &=& - \Omega(T, \mu).  \label{Pr}
\end{eqnarray}
In previous sections, we have estimated all parameters of the two fields and of the gluonic potential. Thus, we can substitute these in Eqs. (\ref{newpotential}) and (\ref{Pr}). Then, the pressure density reads
\begin{eqnarray} \label{Pq}
p(q,\bar{q})&=& T d_q \int_0^{\infty}k^2\, dk \left[\frac{\ln (1-f_q^-(T,\mu ))+\ln (1-f_q^+(T,\mu ))}{\pi ^2} \right. \nonumber \\ 
&& \left. \hspace*{24mm} + \frac{\ln (1-f_s^{-}(T,\mu ))+\ln (1-f_s^{+}(T,\mu ))}{2 \pi ^2}\right],\\
p_g &=& 3\, \pi ^2\, d_g \int_0^{\infty}k^4\, dk \left[ \left(e^{\frac{\omega _g(T,\mu )}{T}}-1\right) \omega _g(T,\mu )\right]^{-1},
\end{eqnarray}
where,
\begin{eqnarray} 
f_q^{\pm} (T,\mu) &=& \frac{1}{e^{\frac{\sqrt{\frac{1}{4} g^2 \sigma_q(T,\mu ){}^2+k^2}\pm \mu }{T}}+1}, \label{var1}\\
f_s^{\pm} (T,\mu) &=& \frac{1}{e^{\frac{\sqrt{\frac{1}{2} g^2 \sigma_s(T,\mu ){}^2+k^2}\pm \mu }{T}}+1}, \label{var2}\\ 
\omega _g(T,\mu ) &=& \sqrt{k^2+\frac{8 \pi ^2 \left(\frac{9 \mu }{2 \pi ^2}+\left(\frac{N_f}{2}+3\right) T^2\right) \left(1-\frac{3 \left(34 N_c^2-13 N_c \, N_f+\frac{3 N_f}{N_c}\right) \ln \left(\ln ^2\left(\frac{\lambda  (T-T_s)}{T_c}\right)\right)}{(11 N_c-2 N_f)^2 \ln ^2\left(\frac{\lambda  (T-T_s)}{T_c}\right)}\right)}{(11 N_c-2 N_f) \ln ^2\left(\frac{\lambda  (T-T_s)}{T_c}\right)}}. \hspace*{10mm} \label{var3}
\end{eqnarray}

\begin{figure}[htb]
\centering{
\includegraphics[width=7.cm,angle=-90]{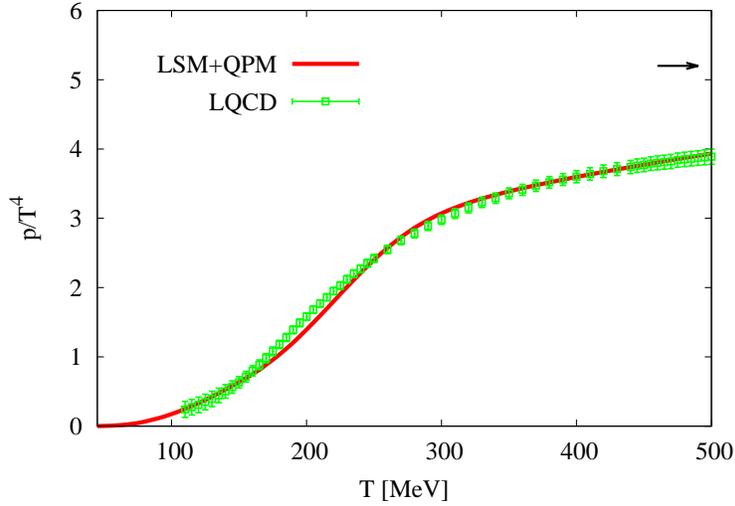}
\caption{The thermal behavior of pressure calculated in the present approach (solid curve) is compared with the lattice QCD pressure (dotted curve) \cite{QCDL} at vanishing chemical potential.  \label{fig:pr} }
}
\end{figure}

In Fig. \ref{fig:pr}, the thermal behavior of the pressure density calculated in LSM with contributions from quasi-gluons (QPM) is compared with recent lattice QCD calculations at vanishing chemical potential \cite{QCDL}. Below $T_c$ LSM is excellently able to reproduce the lattice QCD. Any apparent discrepancy with Fig. \ref{fig:prold} might be understood due to $T_c$-differences. Above $T_c$, we take into consideration contributions from the quasi-particle model (gluonic sector). In fact, it could be included also below $T_c$. At $T< T_c$, the QPM contribution to the thermodynamic quantities calculated in LSM is small. At $T\geq T_c$, the gluonic sector comes up with a considerable contribution. The latter practically represents the motivation of this present approach. We observe that the pressure increases with increasing $T$ until it gets relative close to the value of massless gas (Stefan-Boltzmann limit). Below $T_c$, the pressure is calculated using the quark potential (LSM), only. Above $T_c$, we add the quasigluonic potential part and use $g=8.3$, $\lambda=2.0 $, $T_s=0.0~$MeV and quark masses at zero temperature $m_q=500 ~$MeV and $m_s=1075~$MeV. The agreement seems to be satisfactorily good.

\item{\it The entropy.}\\
At constant volume and vanishing chemical potential, the entropy density is given by the derivative of effective potential with respect to the temperature, 
 \begin{eqnarray} \label{en}
s =  \dfrac{\partial p}{\partial T}.
\end{eqnarray}
By using numerical differentiation of pressure, we get the entropy density
\begin{eqnarray} \label{Sq}
s(q,\bar{q}) & = & \int_0^{\infty} dk \left[\dfrac{P(q,\bar{q})}{T} +\frac{T k^2 d_q}{2 \pi ^2}  \left(-\frac{2 f_q^{-(1,0)}(T,\mu )}{f_q^{-}(T,\mu )-1}-\frac{2 f_q^{+(1,0)}(T,\mu )}{f_q^{+}(T,\mu )-1}\right. \right. \nonumber \\ 
& & \left.
\left.\left.+\frac{f_s^{-(1,0)}(T,\mu )}{1-f_s^{-}(T,\mu )}+\frac{f_s^{+(1,0)}(T,\mu )}{1-f_s^{+}(T,\mu )}\right)\right.\right],\\
s_g & = & \frac{d_g}{3 \pi ^2 T^2} \int_0^{\infty}k^4\, dk \left[\frac{T^2 \omega_g^{(1,0)}(T,\mu )+e^{\frac{\omega _g(T,\mu )}{T}} \omega_g(T,\mu)^2}{ \left(e^{\frac{\omega_g(T,\mu )}{T}}-1\right){}^2 \omega _g(T,\mu ){}^2} \right. \nonumber \\ && \left.
-\frac{T e^{\frac{\omega_g(T,\mu )}{T}} \left(\omega _g(T,\mu )+T\right) \omega _g{}^{(1,0)}(T,\mu )}{\left(e^{\frac{\omega_g(T,\mu )}{T}}-1\right)^2 \omega_g(T,\mu)^2}\right].
\end{eqnarray}
From Eqs. (\ref{var1}) - (\ref{var3}), we get
\begin{eqnarray} 
f_q^{\pm (1,0)}(T,\mu)&=& \dfrac{\partial f_q^{\pm}}{\partial T}, \label{var12}\\
f_s^{\pm (1,0)}(T,\mu)&=& \dfrac{\partial f_s^{\pm}}{\partial T}, \label{var22}\\
\omega_g{}^{(1,0)}(T,\mu )&=& \dfrac{\partial \omega _g(T,\mu )}{\partial T}. \label{var32}
\end{eqnarray}

\begin{figure}[htb]
\centering{
\includegraphics[width=7.cm,angle=-90]{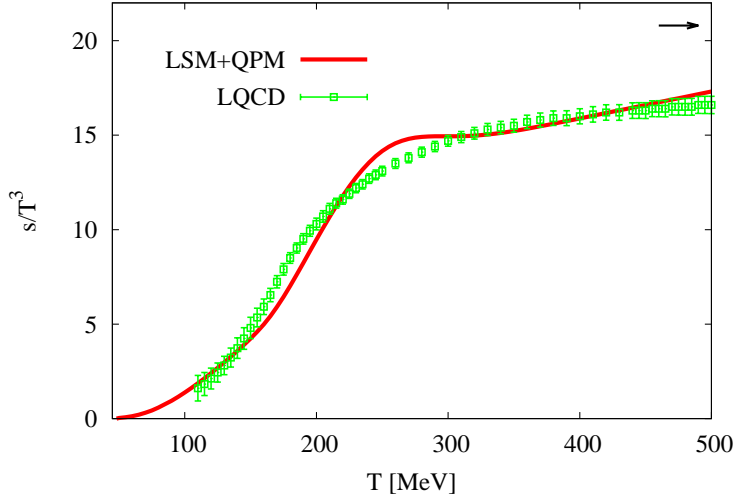}
\caption{The thermal behavior of entropy density from the present approach (solid curve) is compared with the most-recent lattice QCD calculations (dotted curve) at vanishing chemical potential \cite{QCDL}. 
\label{fig:S}}
}
\end{figure}

In Fig. \ref{fig:S}, the thermal behavior of the entropy density calculated in LSM below $T_c$ and LSM plus QPM above $T_c$ is compared with recent lattice QCD calculations at vanishing chemical potential \cite{QCDL}. It is obvious that the entropy density increases with increasing $T$ until it gets close to the value of massless gas (Stefan-Boltzmann limit). The agreement is obviously convincing.  The peak around $T_c$ is an artifact of the heavy masses.

\item{\it The energy density. }\\
To complete the thermodynamic comparison, we also show the energy density $\epsilon$,
\begin{eqnarray} \label{enr}
\epsilon = -p + T\, s.
\end{eqnarray}
By using Eqs. (\ref{eq:quark_pot}), (\ref{Upotio}), (\ref{Eq:Ug}) and (\ref{en}), we get a full expression for the energy density
\begin{eqnarray} \label{Eq}
\varepsilon(q,\bar{q})&=&\frac{T^2 d_q}{2\, \pi^2} \int_0^{\infty} k^2 \, dk \left\{-\frac{2 f_q^{-(1,0)}(T,\mu )}{f_q^{-}(T,\mu )-1}-\frac{2 f_q^{+(1,0)}(T,\mu )}{f_q^{+}(T,\mu )-1} + \frac{f_s^{-(1,0)}(T,\mu )}{1-f_s^{-}(T,\mu )}+\frac{f_s^{+(1,0)}(T,\mu )}{1-f_s^{+}(T,\mu )}\right\}, \hspace*{10mm}\\
\varepsilon_g &=& \frac{d_q}{3\, \pi^2} \int_0^{\infty}k^4\, dk \left\{\frac{\omega _g(T,\mu )+T \omega _g{}^{(1,0)}(T,\mu )}{\left(e^{\frac{\omega_g(T,\mu )}{T}}-1\right)^2 \omega_g(T,\mu ){}^2}  \right. \nonumber \\ 
&& \hspace*{25mm} \left. - \frac{e^{\frac{\omega _g(T,\mu )}{T}} \left[\left(T-\omega _g(T,\mu )\right) \omega _g(T,\mu )+T \left(\omega _g(T,\mu )+T\right) \omega _g^{(1,0)}(T,\mu )\right]}{T \left(e^{\frac{\omega _g(T,\mu )}{T}}-1\right)^2 \omega _g(T,\mu )^2}\right\}.
\end{eqnarray}

\begin{figure}[htb]
\centering{
\includegraphics[width=7.cm,angle=-90]{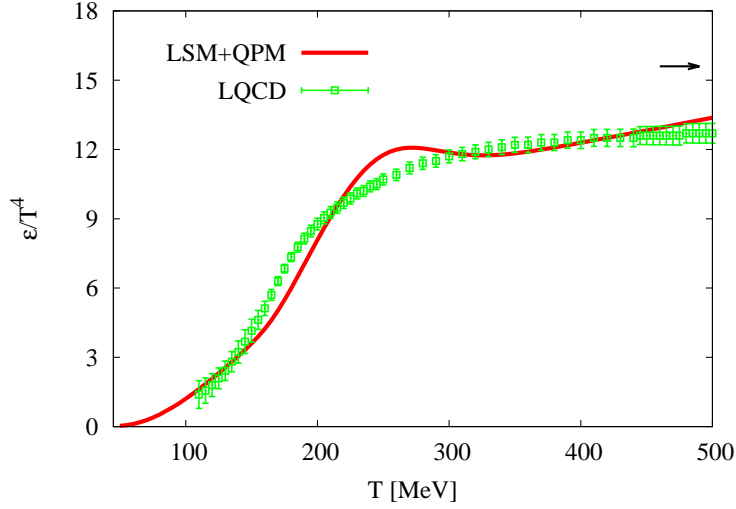}
\caption{The thermal dependence of the energy density calculated in the present approach (solid curve) is compared with the lattice QCD  (dotted curve) \cite{QCDL} at vanishing chemical potential. \label{fig:En}}
} 
\end{figure}

In Fig. \ref{fig:En}, the behavior of the energy density calculated in LSM below $T_c$ and LSM plus QPM (bove $T_c$) is compared with the lattice QCD calculations at vanishing chemical potential \cite{QCDL}.  In other words, we present a comprehensive comparison with recent lattice QCD \cite{QCDL} at vanishing chemical potential. The normalized energy density increases with $T$ until it gets close to the value of massless gas (Stefan-Boltzmann limit). The agreement is excellent. 

\end{itemize}

\subsection{Number density and Susceptibility}
\label{sec:higher}

In this section, we study dimensionless first and second order moments of the particle multiplicity \cite{Tawfik:2014uka}. This can be done with respect to the chemical potential $\mu$ (or temperature $T$) \cite{Tawfik:2014uka}. Here, we implement the approach that the chiral phase transition should be accompanied with large fluctuations in dimensionless quantities,  such as susceptibility. Using Eq. (\ref{newpotential}) we get
\bea \label{Eq:n}
n(T,\mu)=\dfrac{\partial p(T,\mu)}{\partial \mu} &=& \dfrac{\partial p_g}{\partial \mu} + \dfrac{\partial p(q,\bar{q})}{\partial \mu}, 
\eea
where $\partial P_g/\partial \mu$ ($\partial P(q,\bar{q})/\partial \mu$) being gluonic (quark-antiquark) number density.  
\begin{eqnarray} \label{nq}
n(q,\bar{q})&=& -T \frac{d_q}{2 \pi ^2} \int_0^{\infty}k^2\, dk \left[\frac{2 f_q^{-(0,1)}(T,\mu )}{f_q^{-}(T,\mu )-1}+\frac{2 f_q^{+(0,1)}(T,\mu )}{f_q^{+}(T,\mu )-1}+\frac{f_s^{-(0,1)}(T,\mu )}{f_s^{-}(T,\mu )-1}+\frac{f_s^{+(0,1)}(T,\mu )}{f_s^{+}(T,\mu )-1}\right] \hspace*{10mm}\\
n _g&=&\frac{d_g}{3 \pi ^2 T}\int_0^{\infty}k^4\, dk \frac{ \left[T - e^{\frac{\omega _g(T,\mu )}{T}} \left(\omega _g(T,\mu )+T\right)\right] \omega _g{}^{(0,1)}(T,\mu )}{ \left(e^{\frac{\omega _g(T,\mu )}{T}}-1\right)^2 \omega _g(T,\mu )^2},
\end{eqnarray}
\begin{eqnarray} 
f_q^{\pm (0,1)}(T,\mu)&=& \dfrac{\partial f_q^{\pm}}{\partial \mu}, \label{var13}\\
f_s^{\pm (0,1)}(T,\mu)&=& \dfrac{\partial f_s^{\pm}}{\partial \mu}, \label{var23}\\
\omega_g^{(0,1)}(T,\mu )&=& \dfrac{\partial \omega_g(T,\mu )}{\partial \mu}. \label{var33}
\end{eqnarray}

The susceptibility reads
\bea \label{Eq:chi}
\chi(T,\mu)=\dfrac{\partial n(T,\mu)}{\partial \mu} &=& \dfrac{\partial n_{g}}{\partial \mu} + \dfrac{\partial n_{q \bar{q}}}{\partial \mu},
\eea 
where $\partial n_{g}/\partial \mu$ ($\partial n_{q \bar{q}}/\partial \mu$) being gluonic (quark-antiquark) susceptibility
\begin{eqnarray} \label{chi}
\chi(q,\bar{q})&=&\frac{d_q}{2 \pi ^2} T \int_0^{\infty}k^2\, dk \left[\frac{2 f_q^{-(0,1)}(T,\mu )^2-2 (f_q^{-}(T,\mu )-1) f_q^{-(0,2)}(T,\mu )}{ (f_q^{-}(T,\mu )-1)^2}  \right.  \nonumber \\ 
&& \left. \hspace*{27mm} 
+\frac{2 f_q^{+(0,1)}(T,\mu )^2-2 (f_q^{+}(T,\mu )-1) f_q^{+(0,2)}(T,\mu )}{ (f_q^{+}(T,\mu )-1)^2}  \right. \nonumber \\ 
&& \left. \hspace*{27mm} 
+\frac{f_s^{-(0,1)}(T,\mu )^2-(f_s^{-}(T,\mu )-1) f_s^{-(0,2)}(T,\mu )}{(f_s^{-}(T,\mu )-1)^2}    \right. \nonumber \\ 
&& \left. \hspace*{27mm} 
+\frac{f_s^{+(0,1)}(T,\mu )^2-(f_s^{+}(T,\mu )-1) f_s^{+(0,2)}(T,\mu )}{(f_s^{+}(T,\mu )-1)^2}\right],\\
\chi _g &=&\frac{d_g}{3 \pi ^2 T^2 } \int_0^{\infty} \frac{k^4\, dk }{\left(e^{\frac{\omega _g(T,\mu )}{T}}-1\right){}^3 \omega _g(T,\mu ){}^3} \left[2 T^2 \omega _g{}^{(0,1)}(T,\mu )^2 \left(e^{\frac{\omega _g(T,\mu )}{T}}-1\right){}^2\right.    \nonumber \\ &&
- 2 T e^{\frac{\omega _g(T,\mu )}{T}} \omega _g(T,\mu ) \omega _g^{(0,1)}(T,\mu )^2+e^{\frac{2 \omega _g(T,\mu )}{T}} \omega _g(T,\mu ) \left(\omega _g(T,\mu )+2 T\right) \omega _g{}^{(0,1)}(T,\mu )^2 \nonumber \\ &&
+ \left.  T \left(1-e^{\frac{\omega _g(T,\mu )}{T}}\right) \omega _g(T,\mu ) \left(e^{\frac{\omega _g(T,\mu )}{T}} \left(\omega _g(T,\mu )+T\right)-T\right) \omega _g{}^{(0,2)}(T,\mu )\right],
\end{eqnarray}
where
\begin{eqnarray} 
f_q^{\pm (0,2)}(T,\mu)&=& \dfrac{\partial^2 f_q^{\pm}}{\partial \mu^2}, \label{var14}\\
f_s^{\pm (0,2)}(T,\mu)&=& \dfrac{\partial^2 f_s^{\pm}}{\partial \mu^2}, \label{var24}\\
\omega _g^{(0,2)}(T,\mu )&=& \dfrac{\partial^2 \omega_g(T,\mu )}{\partial \mu^2}. \label{var34}
\end{eqnarray}

\begin{figure}[htb]
\centering{
\includegraphics[width=7.cm,angle=-90]{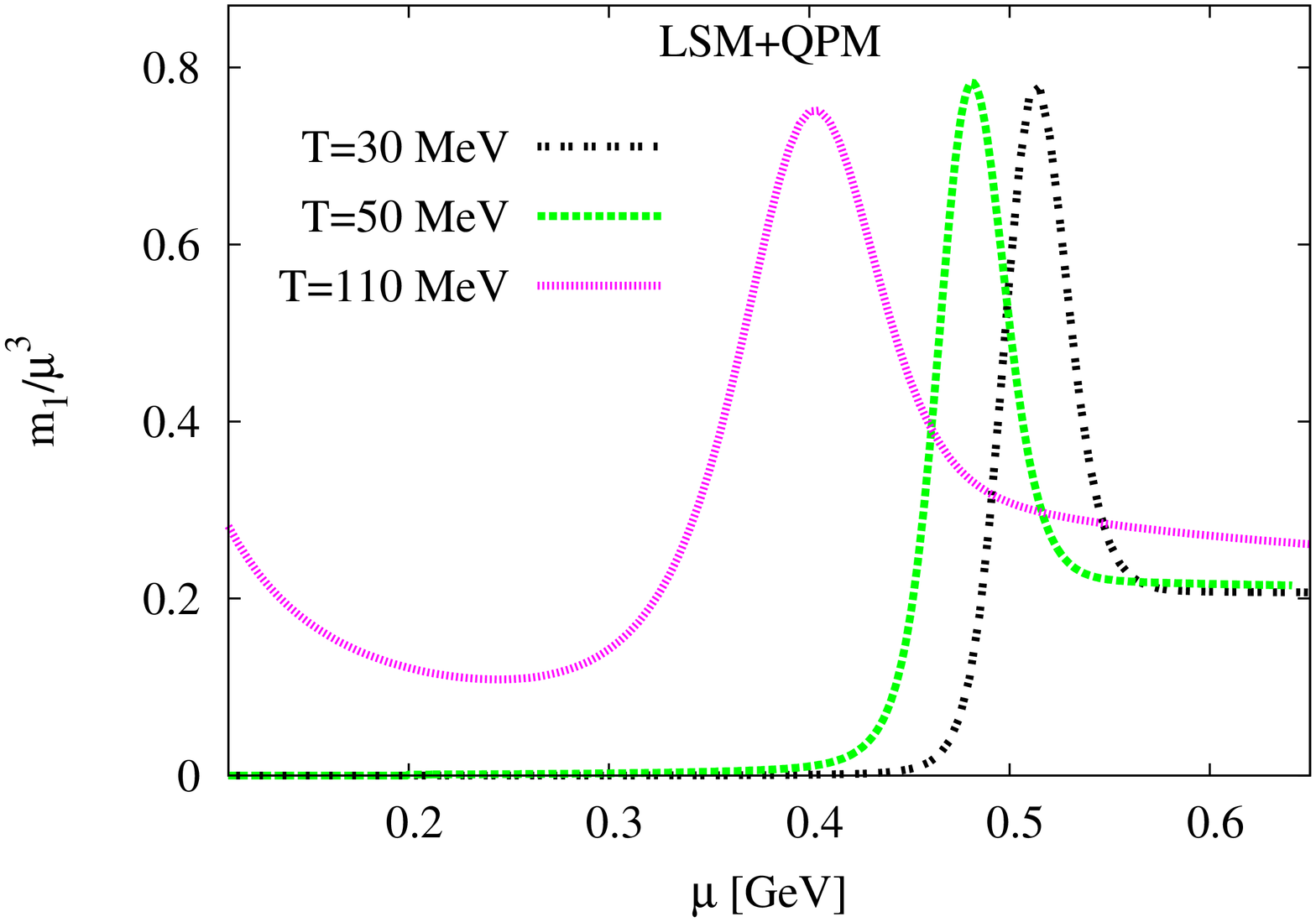}\\
\includegraphics[width=7.cm,angle=-90]{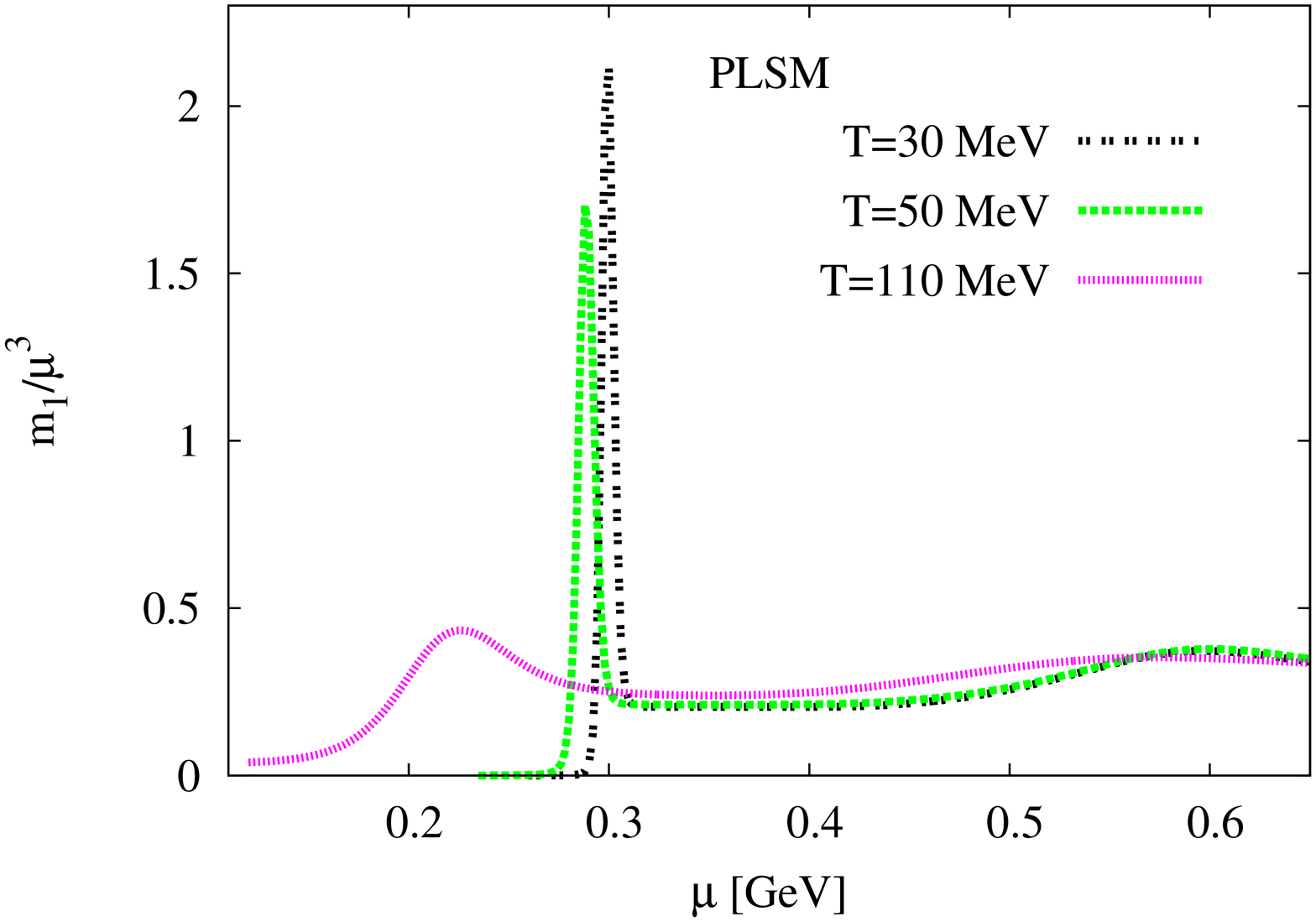}
\caption{Top panel: the dimensionless first order moment is given as a function of the chemical potential at fixed temperatures $30~$MeV (double-dotted curve), $50~$MeV (dashed curve) and $110~$MeV (dotted curve). Bottom panel shows the same as top panel but here results from PLSM \cite{Tawfik:2014uka}.
\label{fig:m1}}
}
\end{figure}

\begin{figure}[htb]
\centering{
\includegraphics[width=7.cm,angle=-90]{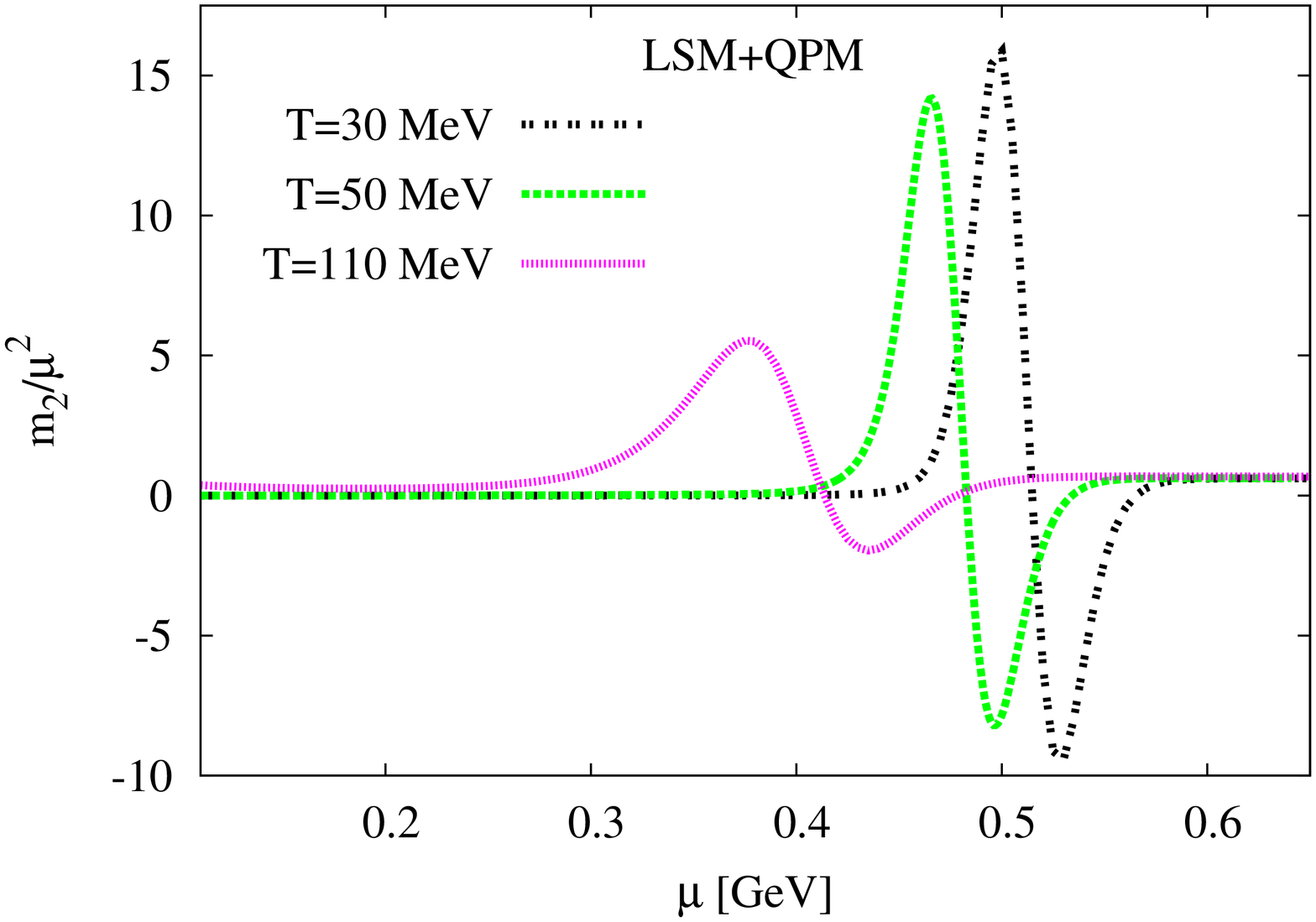}\\
\includegraphics[width=7.cm,angle=-90]{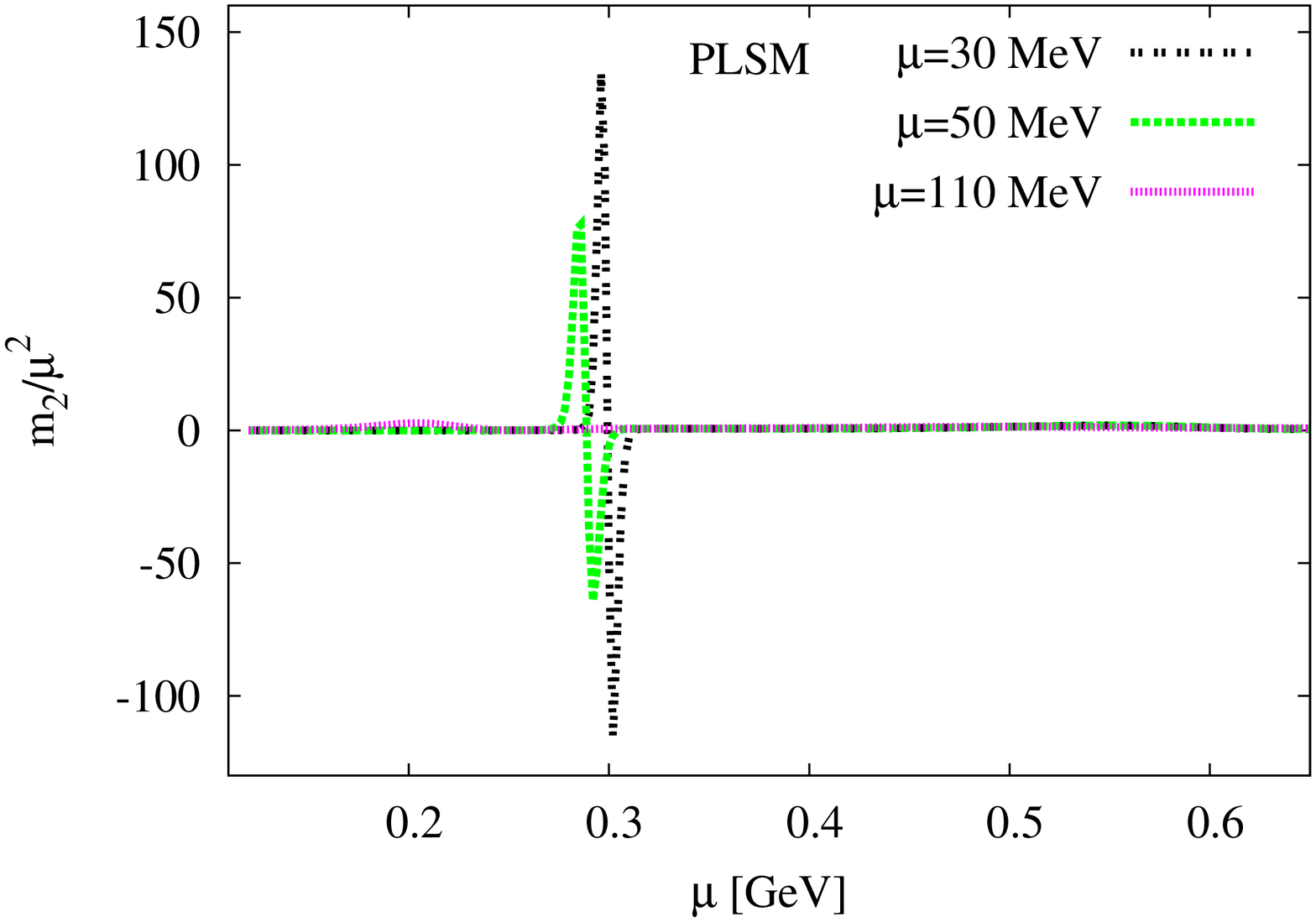}
\caption{The same as in Fig. \ref{fig:m1}, but here for dimensionless second order moment calculated in the present model (top panel) and PLSM (bottom panel) \cite{Tawfik:2014uka}.
\label{fig:m12}}
}
\end{figure}

As given in Eq. (\ref{nq}) the first order moment, $n$, can be normalized to $T^3$ or to $\mu^3$ \cite{Tawfik:2014uka}. In Fig. \ref{fig:m1}, the dimensionless first moment is given as a function of the chemical potential at fixed temperatures: $T=30~$MeV (double-dotted curve), $T=50~$MeV (dashed curve) and $T=110~$MeV (dotted curve). We find that the height of corresponding peak decreases with increasing temperature. Also, the position of the peak moves, when lowering the temperature. In bottom panel, the dimensionless second order moment is given as a function of the chemical potential at the same values of temperatures. Again, the heights of the peaks decrease with increasing temperature and their location on the $\mu$-axis moves with lowering the temperature. The present results are compared with PLSM  \cite{Tawfik:2014uka}, in which Polyakov loop corrections are included and is assumed to well reproduce the lattice QCD calculations \cite{HotQCD}.

From Figs. \ref{fig:m1} and \ref{fig:m12}, we conclude that PLSM and LSM$+$QPM give almost compatible structures of the $\mu$-dependence of first and second order moments, apart from the shift in $\mu$-axis. The present model LSM$+$QPM prefers lower $\mu$-values relative to PLSM.

\subsection{Chiral phase-transition}
\label{sec:chiralpt}

\begin{figure}[htb]
\centering{
\includegraphics[width=7.cm,angle=-90]{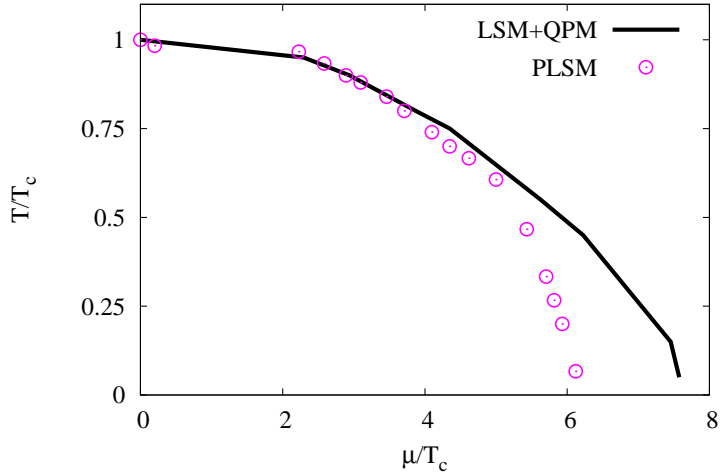}
\caption{The chiral boundary: corresponding temperature and chemical potential are normalized to $T_c=195~$MeV for LSM$+$QPM and $T_c=150~$MeV for PLSM. \label{fig:phase}}
}
\end{figure}

The higher order moments of particle multiplicity are assumed to highlight various types of fluctuations in $T$ and $\mu$ dimensions. They likely appear in drastic change in the degrees of freedom, symmetry change/restoration or the dynamics deriving the system out of equilibrium \cite{critPT}.  In the present work, we utilize the possible fluctuations accompanying normalized second order moment \cite{Schaefer:2008hk,Karsch2009a,Tawfik:2014uka} in mapping out the chiral phase-transition. The problematic of determining quasi-critical temperature from the second moment has been discussed  \cite{Karsch2009a}. Accordingly, we observe that the peaks corresponding to different temperatures are conjectured to be characterized by $\mu$-values, where chiral phase-transition is conjectured to take place.   In top panel of Fig. \ref{fig:m12}, we found that $\mu$-values of the peaks vary with $T$. We scan this dependence at different values of $T$. Then, we follow the scheme to determine $T$ and $\mu$, at which $\chi/\mu^2$ gets maximum. The results are given in Fig. \ref{fig:phase} and compared with PLSM \cite{Tawfik:2014uka}. In doing this, we normalize both quantities to the corresponding critical temperature. The two configurations of SU(3) $\sigma$-model are non-distinguishable, especially at low chemical potential. As given in Ref. \cite{Tawfik:2014uka}, the lattice and experimental freeze-out parameters fit well with PLSM in this limit. However, we note that the chiral temperature of LSM$+$QPM gets larger than that of  PLSM at large chemical potential.

\section{Conclusions and outlook}
\label{sec:conclusion}

In the present work, we started with LSM, but added a quasi-particle sector, especially at temperatures larger than $T_c$. This is an original approach, especially in light of the excellent agreement with recent lattice QCD simulations \cite{QCDL}. Dealing with LSM$+$QPM is also new with respect to combining a first-principle and QCD-like approach with an effective one, the quasi-particle model. The latter gives a phenomenological description treating the interacting massless quarks and gluons as non-interacting massive quasi-particles \cite{qusim1}. The corresponding degrees of freedom are treated in a similar way as one treats electrons in condensed matter theory \cite{paul}; the interaction with the medium provides them with a dynamic mass. Consequently, most interactions can be taken into account. Furthermore, the original QPM has three free parameters be fixed through confronting the model with the lattice QCD calculations. We noticed that the resulting LSM$+$QPM has one free parameter.  

Furthermore, we confronted the results with the first-principle lattice QCD calculations in a systematic way and estimated the higher moments of the particle production in temperature- and density-dependence. This is original as well. Finally, we mapped the chiral phase diagram. Again this is an original approach. The comparison between PLSM and LSM$+$QPM shows that both models are able to reproduce recent LQCD \cite{HotQCD} and \cite{QCDL}, respectively, especially at low chemical potential (very high center-of-mass energy).

The recent lattice QCD calculations give almost identical results below $T_c$, in the hadronic phase. Therefore, we conclude that LSM without any corrections or extensions is not able to reproduce the lattice QCD thermodynamics. In a previous work \cite{Tawfik:2014uka}, we find that PLSM reproduces the lattice results obtained by the HotQCD collaboration in the hadronic and partonic phases. In order to simulate recent lattice calculations \cite{QCDL}, we have to add the gluonic section of the QPM to the pure LSM. The latter seems to provide the correct degrees of freedom, effective masses, couplings and the dynamics needed to regenerate the lattice thermodynamics. 

The thermal behavior of all thermodynamic quantities calculated in LSM with contributions from quasi-gluons (QPM) is compared with recent lattice QCD calculations at vanishing chemical potential \cite{QCDL}. Below $T_c$, LSM is excellently able to reproduce the lattice QCD. Above $T_c$, we take into consideration contributions from the quasi-particle model (gluonic sector), in order to reproduce lattice thermodynamics.  We use $g=8.3$, $\lambda=2.0 $, $T_s=0.0~$MeV and quark masses at zero temperature $m_q=500~$MeV and $m_s=1075~$MeV. The agreement seems to be satisfactorily good. The peak around $T_c$ is an artifact of the heavy masses.

The peaks of higher moments corresponding to different temperatures are conjectured to be characterized by $\mu$-values, where chiral phase-transition is conjectured to take place. We found that $\mu$-values of the peaks vary with $T$. We scan this dependence at different values of $T$. Then, we follow the scheme to determine $T$ and $\mu$, at which $\chi/\mu^2$ gets maximum. The two configurations of SU(3) $\sigma$-model are non-distinguishable, especially at low chemical potential. The lattice and experimentally deduced freeze-out parameters fit well with PLSM in this limit. However, we note that the chiral temperature of LSM$+$QPM gets larger than that of  PLSM at large chemical potential.

In intends to confront the new configuration of LSM and QPM to lattice QCD calculations at finite chemical potential. As discussed in section \ref{sec:motv}, reproducing lattice QCD simulations is apparently sensitive to the utilized action. The present work highlights the advantage of adding LSM and QPM since the transition in this model is much more slower than in the original LSM and also than PLSM. Apparently, this slow transition agrees well with the recent lattice QCD calculations \cite{QCDL}.

\section*{Acknowledgement}
The present work was supported by the World Laboratory for Cosmology And Particle Physics (WLCAPP) http://wlcapp.net/



\end{document}